\documentclass[preprint,aps,a4paper,eqsecnum,amsmath,amssymb]{revtex4}

\usepackage{graphicx}

\newcommand{\id}{I}
\newcommand{\I}{\mathrm{i}}
\newcommand{\etal}{\emph{et al.~}}
\newcommand{\TQ}{\ensuremath{T}\ensuremath{Q}-}
\newcommand{\ch}{\cosh}  
\newcommand{\sh}{\sinh}  
\newcommand{\End}{\operatorname{End}}   
\newcommand{\tr}{\operatorname{tr}} 
\newcommand{\tnh}{\tanh}
\DeclareMathOperator{\diag}{diag}
\newcommand{\dv}{\mathrm{d}}
\newcommand{\e}{\mathrm{e}}
\newcommand{\bra}[1]{\langle #1\vert}
\newcommand{\ket}[1]{\vert #1\rangle}
\newcommand{\comm}[2]{[#1,#2]}

\begin{document}
\preprint{}
\title{The XXZ model with anti-periodic twisted boundary conditions}

\author{S\"onke Niekamp}
\altaffiliation[Present Address: ]{Institut f\"ur Quantenoptik und
  Quanteninformation, Technikerstra\ss e 21a, A-6020 Innsbruck, Austria.}

\author{Tobias Wirth}

\author{Holger Frahm}

\affiliation{%
Institut f\"ur Theoretische Physik, Leibniz Universit\"at Hannover,
Appelstra\ss{}e 2, 30167 Hannover, Germany}

\date{March 23, 2009} 

\begin{abstract}
  We derive functional equations for the eigenvalues of the XXZ model subject
  to anti-diagonal twisted boundary conditions by means of fusion of transfer
  matrices and by Sklyanin's method of separation of variables.  Our findings
  coincide with those obtained using Baxter's method and are compared to the
  recent solution of Galleas.  As an application we study the finite size
  scaling of the ground-state energy of the model in the critical regime.
\end{abstract}

\maketitle
\section{Introduction}
The Quantum Inverse Scattering Method (QISM) \cite{vladb} provides a powerful
framework for the construction of exactly solvable lattice models with the
Yang-Baxter equation defining the underlying algebraic structure.  This
structure also serves as the basis for the solution of the spectral problem,
i.e.\ explicit construction of eigenvalues and eigenstates, for the transfer
matrices of these lattice models, by means of the algebraic Bethe ansatz.  The
application of this method, however, is limited to systems with a simple
pseudo vacuum state which can be used as a reference state to generate the
complete spectrum by application of the elements of the Yang-Baxter algebra.
There exist methods to compute the spectrum of these systems which do not
suffer these restrictions, e.g.\ Baxter's method of commuting transfer
matrices \cite{baxter:book} or Sklyanin's method for separation of variables
\cite{Skly92}.  In these approaches, the eigenvalues are encoded into the
solutions to certain functional relations which have to be solved in a second
step: using the analytical properties of the transfer matrix its eigenvalues
can be parametrized in terms of the roots of Bethe equations.

Recently, such alternative routes to the solution of the spectral problem of
exactly solvable models have been applied to anisotropic spin chains with open
ends and general non-diagonal boundary fields.  Here, the boundary fields
break particle number conservation and therefore the completely polarized
state of spins cannot be used as a pseudo vacuum for the algebraic Bethe
ansatz.  Various methods have been used to determine the spectrum of the
corresponding transfer matrix
\cite{Nepo04,MeRM05,MuNS06,YaNZ06,Galleas08,FrSW08,BaKo07}.  Some of these methods
require restrictions on the boundary fields and/or the bulk anisotropy of the
spin chain to work, other approaches rely on conjectures regarding the
existence of certain limits.  Depending on the specific approach one obtains
very different functional equations for the spectral problem and in many cases
an efficient procedure for their solution is still missing.

In this paper we consider some of these approaches in a simpler setting to
better understand how they are related and to assess their applicability for
the solution of the spectral problem: the XXZ model with general toroidal
boundary conditions is defined by the hamiltonian
\begin{equation}
\label{hamil}
  \mathcal{H} = \sum_{j=1}^{L}\Big[\sigma_j^x\sigma_{j+1}^x +
  \sigma_j^y\sigma_{j+1}^y + \cosh\eta\,\sigma_j^z\sigma_{j+1}^z \Big]\,, \qquad
  \sigma^\alpha_{L+1}=K^{-1} \sigma^\alpha_1 K 
\end{equation}
where $\sigma_j^\alpha$, $\alpha=x,y,z$ denote the Pauli matrices for spins
$\frac{1}{2}$ at site $j$.  The unitary matrix $K \in \End(\mathbb{C}^2)$ determines
the boundary conditions.  For anti-diagonal $K$ the model is integrable but
has no pseudo vacuum state (see below).  It has first been solved by means of
Baxter's method \cite{BBOY95} and solutions to the resulting functional
equations can be given in terms of the roots of Bethe ansatz equations.

The integrability of (\ref{hamil}) is established within the QISM
\cite{vladb}
from the Yang-Baxter algebra 
\begin{equation}\label{YBA}
R_{12}(\lambda-\mu) T_1(\lambda) T_2(\mu) = T_2(\mu) T_1(\lambda) R_{12}(\lambda-\mu)\,.
\end{equation}
Here the monodromy matrix $T_j(\lambda)$ is a matrix on the auxiliary linear
space $V_j$ with entries being the generators of the quadratic algebra.  The
structure constants are arranged in the $R$-matrix, $R_{jk}(\lambda)
\in \End(V_j\otimes V_k)$ which itself solves the Yang-Baxter equation
\begin{equation} \label{YBE}
R_{12}(\lambda-\mu)\,R_{13}(\lambda-\nu)\,R_{23}(\mu-\nu)
= R_{23}(\mu-\nu)\,R_{13}(\lambda-\nu)\,R_{12}(\lambda-\mu)\,.
\end{equation}
In this paper we will use the well-known trigonometric solution for
two-dimensional spaces $V_j$ corresponding to the six-vertex model.  In this
case a representation of (\ref{YBA}) is given by local $\cal L$-matrices with
$T_a =R_{aj}\equiv\mathcal{L}_j$ reading
\begin{equation}
\label{lop}
{\cal L}_j (\lambda) = 
  \begin{pmatrix}  
	\sh (\lambda) S^0_j+ \ch (\lambda) S_j^z &  S_j^- \\
	 S_j^+          & \sh (\lambda) S^0_j- \ch (\lambda) S_j^z
  \end{pmatrix}\,.
\end{equation}
The elements of ${\cal L}_j$ are operators on a two-dimensional quantum state
space of a spin $\tfrac12$, in terms of the Pauli matrices $\sigma_j$ they are
$S^0_j=\ch(\eta/2)\id$, $S_j^z=\sh(\eta/2)\sigma_{j}^{z}$ and $S_j^\pm=
\sh(\eta/2)\ch(\eta/2) \sigma_j^\pm$.
Using the co-multiplication of the Yang-Baxter algebra new representations can
be constructed from (\ref{lop}): the monodromy matrix $T(\lambda)= {\cal
  L}_L(\lambda)\cdots {\cal L}_1(\lambda)$ also satisfies (\ref{YBA}).
Note that ${\cal L}_j(\lambda)$ (and $T(\lambda)$) satisfies
\begin{equation} \label{L-periodicity}
\mathcal{L}_j(\lambda +\I \pi) = -\sigma^z_0 \mathcal{L}_j(\lambda)\sigma^z_0 = -\sigma^z_j \mathcal{L}_j(\lambda)\sigma^z_j
\end{equation}
where the subscript $0$ denotes the auxiliary space.

Apart from these operator valued representations there exist $2\times2$
c-number solutions $K$ to \eqref{YBA}.  Being independent of the spectral
parameter $\lambda$ they satisfy $\big[R(\lambda),K\otimes K\big] = 0$.
For the six-vertex $R$-matrix, this relation has two classes of solutions,
namely diagonal or anti-diagonal twist matrices $K$.  Without loss of
generality \cite{ABGR88,YuBa95a} we restrict our considerations to
\begin{equation} 
\label{Kmatrices}
K=\begin{pmatrix} \e^{-\I \phi} &0 \\0& \e^{\I \phi}  \end{pmatrix}\quad,\quad K=\begin{pmatrix}0&1\\1&0\end{pmatrix} 
\end{equation}
with a twist angle $\phi$ in the diagonal case.

As a consequence of (\ref{YBA}) the transfer matrix $t(\lambda)=
\tr_{0}[KT(\lambda)]$ generates a family of commuting operators,
$[t(\lambda),t(\mu)]=0$.  Therefore the spin chain hamiltonian \eqref{hamil}
which is obtained from
\begin{equation}
  \label{logderiv}
  H=2\sinh\eta\, \left.\frac{\partial \ln t(\lambda)}{\partial\lambda}
                 \right|_{\lambda=\tfrac{\eta}{2}}
  -L\cosh\eta\ 
\end{equation}
is integrable.


In the following section we will first briefly review the existing solutions
of the XXZ model with anti-diagonal twisted boundary conditions followed by
two different approaches to the spectral problem based on (A) the fusion
hierarchy of transfer matrices \cite{BaRe89} at anisotropies $\eta=\I\pi/(p+1)$
with integer $p>1$ and (B) Sklyanin's method of separation of variables
\cite{Skly92}, respectively.  In Section~\ref{sec:fss} we study the spectrum
of small chains to identify the ground-state and extract the critical
properties from the finite size scaling behaviour of the ground-state energy
in the massless regime $\eta=\I\gamma$ with real $\gamma$.

\section{Solution of the spectral problem for anti-diagonal twist}
For a diagonal twist matrix the eigenvalues and eigenstates of the transfer
matrix $t(\lambda)$ can be obtained by means of the algebraic Bethe ansatz
starting from the ferromagnetic so-called pseudo vacuum with polarized spins,
$\langle\sigma_j^z\rangle = 1$ (see e.g. \cite{vladb}).
For the anti-diagonal twist matrix
\begin{equation}\label{twist-matrix}
  K = \begin{pmatrix} 0 & 1 \\  1 &  0 \end{pmatrix}\,,
\end{equation}
however, the total magnetization is not a conserved quantum number.  As a
consequence there is no simple reference state such as the ferromagnetic one
and the algebraic Bethe ansatz cannot be applied.  Instead, a functional
equation (called \TQ equation) for the eigenvalues $\Lambda(\lambda)$ of the
transfer matrix has been obtained using Baxter's method of commuting transfer
matrices \cite{BBOY95,YuBa95a}
\begin{equation} \label{tqe}
  \Lambda(\lambda) q(\lambda) = \sinh^L\left(\lambda+\frac{\eta}{2}\right)
  q(\lambda-\eta) - \sinh^{L}\left(\lambda-\frac{\eta}{2}\right)
  q(\lambda+\eta)\,.
\end{equation}
This difference equation is solved by
\begin{equation}
\label{Q-func0}
  q(\lambda) = \prod_{j=1}^L \sinh\frac{1}{2}(\lambda-\lambda_j)\,.
\end{equation}
As a consequence of the analyticity of the transfer matrix eigenvalues it
follows that the rapidities $\lambda_j$ are different solutions to the Bethe
equations
\begin{equation} \label{bae}
  \frac{\sinh^L(\lambda+\frac{1}{2}\eta)}{\sinh^L(\lambda-\frac{1}{2}\eta)}
  = - \prod_{k\ne j}
  \frac{\sinh\frac{1}{2}(\lambda-\lambda_k+\eta)}{
        \sinh\frac{1}{2}(\lambda-\lambda_k-\eta)}\,,
  \qquad \mathrm{for}\, \lambda \in \left\{\lambda_j\right\}_{j=1}^{L}\,.
\end{equation}
(Note that these equations with an extra phase $(-1)$ and any number $M\le L$
of rapidities $\lambda_j$ determine the spectrum of a staggered six-vertex
model \cite{IkJS08}.  This case, however, cannot be obtained from
(\ref{hamil}) with the twist matrices \eqref{Kmatrices}.)
{}From (\ref{logderiv}) we obtain the corresponding eigenvalue of the spin
chain hamiltonian (\ref{hamil})
\begin{equation} \label{baenergy}
  E(\{\lambda_j\}) = L \cosh\eta 
    + 2\sum_{j=1}^L\frac{\sinh\eta\,\sinh\tfrac{\eta}{2}}{\cosh{\lambda_j}-\cosh\tfrac{\eta}{2}}\,.
\end{equation}

Recently, Galleas \cite{Galleas08} has proposed a different approach to solve
the spectral problem of the spin chain with anti-diagonal twist. From the
Yang-Baxter algebra he derives a closed set of equations for the $L-1$ zeroes
$\lambda_k^{(1)}$ of the eigenvalues $\Lambda(\lambda)$ \emph{and} a second
set of $2(L-1)$ rapidities $\lambda_\ell^{(2)}$ defined to be the zeroes of
the matrix element of a certain element of the Yang-Baxter algebra between the
ferromagnetic state and the eigenstate of the model with twist
\begin{equation}\label{galle}
\begin{aligned}
  \left( \frac{\sinh(\lambda_k^{(1)}+\frac{1}{2}\eta)}{\sinh\eta}\,
         \frac{\sinh(\lambda_k^{(1)}-\frac{1}{2}\eta)}{\sinh\eta} \right)^{L-1}
  &= -\frac{\mathrm{e}^{\frac{2\I\pi r}{L}}}{L}
     \prod_{\ell=1}^{2(L-1)}
     \frac{\sinh(\lambda_\ell^{(2)}-\lambda_k^{(1)})}{\sinh(\lambda_\ell^{(2)}
       - \frac{1}{2}\eta)} \\
  \left( \frac{\sinh(\lambda_\ell^{(2)}+\frac{1}{2}\eta)}{\sinh\eta}\,
         \frac{\sinh(\lambda_\ell^{(2)}-\frac{1}{2}\eta)}{\sinh\eta}
       \right)^{L-1} 
  &= -\frac{\mathrm{e}^{\frac{2\I\pi r}{L}}}{L}
     \left(\prod_{k=1}^{L-1}
       \frac{\sinh(\lambda_k^{(1)}-\lambda_\ell^{(2)})}{\sinh(\lambda_k^{(1)}
       - \frac{1}{2}\eta)} \right)^2\,.
\end{aligned} 
\end{equation}
Here $r=0,\ldots,2L-1$ is a quantum number fixing the translational properties
of the corresponding eigenstate.  Note that as a consequence of the functional
\TQ equation (\ref{tqe}) $\lambda_k^{(1)}$ are the `hole solutions'
$\lambda=\lambda_k^{(1)}\notin \left\{\lambda_j\right\}_{j=1}^{L}$ to the
Bethe equations (\ref{bae}).

The algebraic equations (\ref{galle})
involving two sets of rapidities are vaguely reminiscent of so-called nested
Bethe equations for systems with higher rank symmetry.  In the present case,
however, the two sets of rapidities are redundant in the sense that -- for an
eigenstate with given $r$ -- either set can be
eliminated in favour of the other one.  Also, unlike the string hypothesis for
the Bethe rapidities $\lambda_j$ nothing is known about the structure of the
solutions of Galleas' equations in the thermodynamic limit $L\to\infty$, which
would be a prerequisite for both numerical or analytical studies of large
systems.

\subsection{Fusion hierarchy and truncation identity at roots of unity}\label{FusionTruncation}
%
This method was first developed for the RSOS model by Bazhanov \etal
\cite{BaRe89} and was adapted to spin chains by Nepomechie
e.g. \cite{Nepo02,Nepo03}.  Unfortunately this method only works if the values
of the crossing parameter are chosen to be roots of unity $\eta = \I\pi/(p+1)$.
Nevertheless as in the periodic case \cite{Nepo03} the solution obtained is
valid for arbitrary $\eta$ as it coincides with \eqref{tqe}.

Because of the Yang-Baxter algebra higher spin transfer matrices of the XXZ
spin chain obey a so-called fusion hierarchy, i.e. it is possible to construct
a higher spin transfer matrix out of lower spin transfer matrices directly.  A
transfer matrix using a spin-$j$ auxiliary space is constructed using fused
$\mathcal L$-matrices.

The fused spin-$(j,\tfrac12)$ $\mathcal L$-matrix for $j=\tfrac12,1,\tfrac32,\dots$ is
given by \cite{KuSk82,KuRS81,Nepo03} 
\begin{equation}
\mathcal L_{<1 \cdots 2j>2j+1}(\lambda) = P^+_{1\cdots 2j}\mathcal L_{1,2j+1}(\lambda) \mathcal L_{2,2j+1} (\lambda+\eta) \cdots \mathcal L_{2j,2j+1}(\lambda+(2j-1)\eta) P^+_{1 \cdots 2j} \quad,
\end{equation}
where $P^+$ is the projector defined by the sum over all permutation operators
for $n$ spin-$\tfrac12$ spaces 
\begin{equation}
P^{+}_{1\cdots n} = \frac{1}{n!} \sum_\sigma P_\sigma \quad .
\end{equation}
In the same way the fusion of the twist matrix \eqref{twist-matrix} is carried
out yielding the spin-$j$ representation
\begin{equation}
K_{<1 \cdots 2j>} = P^+_{1\cdots 2j}K_{1} K_{2} \cdots K_{2j} P^+_{1 \cdots 2j} \quad .
\end{equation}
The fused monodromy matrix for the chain then reads for $L$ lattice
sites of the original hamiltonian \eqref{hamil}
\begin{equation}
T_{<1 \cdots 2j>}(\lambda) =  \mathcal L_{<1 \cdots 2j>,L}(\lambda) \cdots \mathcal L_{<1 \cdots 2j>,1}(\lambda)
\end{equation}
and tracing out the auxiliary space gives the associated transfer matrix
\begin{equation}
t^{(j)}(\lambda) = tr_{1\cdots 2j} K_{<1 \cdots 2j>}T_{<1\cdots 2j>} (\lambda) \quad .
\end{equation}
For these transfer matrices the following fusion hierarchy with
$j=\tfrac12,1,\tfrac32,\dots$ holds \cite{KuSk82,KiRe86,KiRe87,Nepo03}
\begin{equation}\label{hierachy}
t^{(j+1/2)} (\lambda) = t^{(j)}(\lambda) t^{(1/2)}(\lambda+2j\eta)  - (d_q T)(\lambda+(2j-1)\eta) t^{(j-1/2)}(\lambda) \quad ,
\end{equation}
where $t^{(1/2)}(\lambda) \equiv t(\lambda)$ and $t^{(0)}(\lambda) \equiv \id$.

On the other hand fused transfer matrices can be constructed using
quantum-group theory \cite{KuRe83a, Nepo02}.  The $\mathcal L$-matrices of
higher auxiliary spins from quantum-group constructions have a simple direct
relation to an $\mathcal L$-matrix with lower auxiliary spin at roots of unity
$\eta=\I\pi / (p+1)$ with $p$ being an integer number.  It is also possible to
relate the quantum-group $\mathcal L$-matrices to those constructed via
fusion.  Resulting in the identity at roots of unity
\begin{multline}\label{L-trunc}
B_{1\cdots p+1} A_{1\cdots p+1} \mathcal L_{<1\cdots p+1>,p+2}(\lambda) A^{-1}_{1\cdots p+1} B^{-1}_{1\cdots p+1} =\\
=\mu(\lambda)
\begin{pmatrix}
\nu(\lambda)\sigma^z &  & \\
& B_{1\cdots p-1} A_{1\cdots p-1} \mathcal L_{<1\cdots p-1>,p}(\lambda+\eta) A^{-1}_{1\cdots p-1} B^{-1}_{1\cdots p-1} &\\
& & -\nu(\lambda) \sigma^z \\
\end{pmatrix}
\end{multline}
where the entries of matrix $A$ are unnormalized Clebsch-Gordon coefficients
in the decomposition of the tensor product of $2j$ spin-$1/2$ representations
into a direct sum of SU(2) irreducible representations and the matrix $B$ is a
diagonal matrix needed for symmetrizing \cite{Nepo02}.

The function $\mu(\lambda)$ is related to the quantum determinant
\begin{align}
(d_q T)(\lambda - \eta) = - \left( -\mu(\lambda) \right)^L\\
\mu(\lambda) \equiv - \sh(\lambda+\tfrac12 \eta) \sh(\lambda-\tfrac32 \eta)
\end{align} 
and $\nu (\lambda) \equiv - \mu(\lambda)^{-1} (\I/2)^p \sh\left(
  (p+1)(\lambda-\tfrac12\eta) \right)$ is related to the crossing parameter via
$p$.

The fused twist matrices themselves obey a truncation identity similar to
\eqref{L-trunc}.  Under the fusion procedure an anti-diagonal matrix with only
1's as entries remains anti-diagonal with 1's as entries after applying the
transformation of the appropriate Clebsch-Gordon matrix and omitting null rows
and columns, hence
\begin{equation} \label{K-trunc}
A_{1\cdots 2j} K_{<1\cdots 2j>} A^{-1}_{1\cdots 2j} = 
\begin{pmatrix}
&&1 \\
& A_{1\cdots 2j-2} K_{<1\cdots 2j-2>} A^{-1}_{1\cdots 2j-2}&\\
1&&\\
\end{pmatrix} \quad .
\end{equation}

Identities \eqref{L-trunc} and \eqref{K-trunc} together give a truncation
identity for the product of twist and monodromy matrix
\begin{multline}\label{T-trunc}
B_{1\cdots p+1} A_{1\cdots p+1} K T_{<1\cdots p+1>,p+2}(\lambda) A^{-1}_{1\cdots p+1} B^{-1}_{1\cdots p+1} =\\
= \mu(\lambda)^L
\begin{pmatrix}
&  & \left( -\nu(\lambda)\right)^L F \\
& B_{1\cdots p-1} A_{1\cdots p-1} K T_{<1\cdots p-1>,p}(\lambda+\eta) A^{-1}_{1\cdots p-1} B^{-1}_{1\cdots p-1} &\\
\nu(\lambda)^L F \\
\end{pmatrix}
\end{multline}
with $F\equiv \prod_{j=1}^L \sigma_j^z$, and accordingly for the transfer matrix by taking the trace of \eqref{T-trunc}
\begin{equation} \label{truncation}
t^{(p+1)/2}(\lambda) = -(d_q T)(\lambda-\eta) (-1)^{L} t^{(p-1)/2}(\lambda+\eta) \quad . 
\end{equation}

The fusion hierarchy \eqref{hierachy} together with the truncation identity
\eqref{truncation} leads to a functional relation for the transfer matrix at
roots of unity for a given $p$, e.g. for $p=2$ or $j=1$ respectively this
relation is
\begin{multline}\label{func-rel-p2}
t(\lambda) t(\lambda+\eta) t(\lambda+2\eta) - (d_q T)(\lambda) t(\lambda+2\eta) +(d_q T)(\lambda+\eta) t(\lambda)+\\
+ (-1)^{L} (d_q T)(\lambda-\eta) t(\lambda+\eta)  = 0\,.
\end{multline}

Like in the RSOS model \cite{BaRe89} or the periodic XXZ chain \cite{Nepo03}
the goal is to recast the general form of the functional relation
\eqref{func-rel-p2} as a determinant of a certain matrix. This determinant
being zero ensures the existence of a null eigenvector which leads to
equations similar to \TQ equations.

In the case of an anti-diagonal $K$-matrix the functional relation found above
cannot be recast directly, though multiplying it with itself shifted by $\I
\pi = (p+1) \eta$ results in a recastable expression.
For general $p$ this is a determinant of a $(2p+2) \times (2p+2)$ matrix
reading with the eigenvalue $\Lambda$ of the transfer matrix $t$
\begin{equation} \label{recastdeterminant}
\det \begin{pmatrix}
\Lambda_0 & h_0       & 0         & \dots   &     0        & - (-1)^N h_1 \\
-h_2      & \Lambda_1 & h_1       &         &              & 0            \\
0         & -h_3      & \Lambda_2 &  \ddots &              & \vdots       \\
\vdots    &           &  \ddots   &  \ddots &              & 0             \\
 0        &           &           &-h_{2p+1}& \Lambda_{2p} & h_{2p}        \\
(-1)^N h_{2p+1} & 0   & \dots     &  0      & -h_0         & \Lambda_{2p+1} \\
\end{pmatrix} = 0 \quad .
\end{equation}
In the above expression we used the shorthands
\begin{align}
\Lambda_k \equiv &\Lambda(\lambda + k \eta) \\
h_k \equiv& \sh^L(\lambda + k \eta - \frac{\eta}{2}) \quad .
\end{align}
The definition of $h_k$ directly reveals $h_k = (-1)^L h_{p+1+k}$. This and
the periodicity of the eigenvalue $\Lambda_k = - (-1)^L \Lambda_{p+1+k}$
following from \eqref{L-periodicity} are needed to verify the equivalence of
the determinant and the product of functional relations.

Let $\left( q_0, q_1, \ldots, q_{2p+1} \right)$ be the null eigenvector of the
matrix, this yields the equations
\begin{align} \label{eigenvec_equations}
\nonumber \Lambda_0 q_0 +h_0 q_1 - (-1)^L h_1 q_{2p+1} &= 0\\
-h_{k+1}q_{k-1}+\Lambda_k q_k + h_k q_{k+1} & = 0 \quad \text{ for } k=1,\dots,2p\\
\nonumber (-1)^L h_{2p+1} q_0 -h_0 q_{2p} +\Lambda_{2p+1} q_{2p+1} &= 0 \quad .
\end{align}

Using the ansatz $q_k = q (\lambda + k\eta)$ with
\begin{equation} \label{Q-Func}
q(\lambda) = \prod_{j=1}^L \sh \frac{1}{2}( \lambda - \lambda_j )
\end{equation}
the equations \eqref{eigenvec_equations} imply only a single \TQ equation
\begin{equation} \label{TQequation}
\Lambda(\lambda) q(\lambda) = \sh^L(\lambda +\tfrac12 \eta) \, q(\lambda-\eta) - \sh^L(\lambda-\tfrac12 \eta) \, q(\lambda+\eta)
\end{equation}
agreeing with \eqref{tqe} and leading to the same Bethe ansatz equations
\eqref{bae}.
Notice the $2\pi\I$ periodicity of the $q$-function arising from the $2(p+1)$
rows of the matrix in \eqref{recastdeterminant} and the product in
\eqref{Q-Func} running up to $L$ due to the structure of the upper right and
lower left entries.

\subsection{Separation of variables}
In this section, we carry out the procedure of separation of variables,
generalizing Sklyanin's result for the XXX chain~\cite{Skly92}.

We modify our definition of the monodromy matrix by introducing inhomogeneities $\delta_{j}$
\begin{equation}
  T(\lambda)={\cal L}_{L}(\lambda-\delta_{L})\cdots{\cal L}_{1}(\lambda-\delta_{1})\ .
\end{equation}
Later we will discuss the limit $\delta_{j}\to 0$. We employ the usual notation for the elements of the twisted monodromy matrix
\begin{equation}
  \begin{pmatrix} A(\lambda) & B(\lambda)\\ C(\lambda) & D(\lambda)\end{pmatrix}=KT(\lambda)\ .
\end{equation}
The Yang-Baxter algebra contains the commutation relation
\begin{equation}
  \comm{B(\lambda)}{B(\mu)}=0\ .
\end{equation}
It is therefore reasonable to assume that there exists a complete set of
$\lambda$-independent eigenvectors $\ket{\ell}$ of $B(\lambda)$. Note that
this does not follow from the commutativity since we have not shown yet that
$B(\lambda)$ is diagonalizable.
By induction in $L$ we show that in the standard basis $D(\lambda)$ is lower
triangular with all diagonal entries equal to zero, and $B(\lambda)$ is lower
triangular with diagonal entries
\begin{equation}
  \sh(\lambda-x_{1}^{\ell})\cdots\sh(\lambda-x_{L}^{\ell})\qquad\text{where}\qquad x_{j}^{\ell}=\delta_{j}\pm\tfrac{\eta}{2}\ .
\end{equation}
This shows that $B(\lambda)$ is indeed diagonalizable, provided the
inhomogeneities $\delta_{j}$ are mutually distinct, and that its eigenvalues
are given by
\begin{equation}
  \label{scalarroots}
  B(\lambda)\ket{\ell}=\sh(\lambda-x_{1}^{\ell})\cdots\sh(\lambda-x_{L}^{\ell})\ket{\ell}\ .
\end{equation}
In other words, we have defined operator-valued zeroes $\hat{x}_{j}$ of
$B(\lambda)$
\begin{equation}
  B(\lambda)=\sh(\lambda-\hat{x}_{1})\cdots\sh(\lambda-\hat{x}_{L})\ ,
\end{equation}
where $\hat{x}_{j}=\diag(x_{j}^{1},\ldots,x_{j}^{2^{L}})$ in the eigenbasis of
$B(\lambda)$. Since the joint spectrum of the operators $\hat{x}_{j}$ is not
degenerate, any eigenvector $\ket{\ell}$ is completely determined by its
eigenvalues $x_{1}^{\ell},\ldots,x_{L}^{\ell}$. We interpret the set of
eigenvalues $(x_{1}^{\ell},\ldots,x_{L}^{\ell})$ of a given eigenvector
$\ket{\ell}$ as a point in $\mathbb{C}^{L}$. Then the Hilbert space of the
spin chain is isomorphic to the space of complex-valued functions on the set
$X\subset\mathbb{C}^{L}$ of these points (for this reason the separation of
variables method is also called \lq functional Bethe ansatz\rq{}). In this
picture the $\hat{x}_{j}$ are the operators of multiplication by the
coordinate functions $x_{j}$ in $\mathbb{C}^{L}$:
\begin{equation}
  \hat{x}_{j}f=x_{j}f\ ,\qquad\bigl(\hat{x}_{j}f\bigr)(x^{\ell})=x_{j}^{\ell}f(x^{\ell})
\end{equation}
for any function $f$ on $X$. In the following, we shall not distinguish between
the operators $\hat{x}_{j}$ and the functions $x_{j}$.

We want to formulate the spectral problem for the twisted transfer matrix in
the diagonal basis of the operators $x_{j}$. To this end, we first define the
\lq conjugated momenta\rq{} $X_{j}^{\pm}$ to the \lq coordinates\rq{} $x_{j}$,
which we obtain from $A(\lambda)$ and $D(\lambda)$ by substituting
$\lambda=x_{j}$ \lq from the left\rq{}
\begin{equation}
  \bra{\ell}X_{j}^{-}\ket{m}=\bra{\ell}A(x^{\ell}_{j})\ket{m}\qquad\text{and}\qquad\bra{\ell}X_{j}^{+                         }\ket{m}=\bra{\ell}D(x^{\ell}_{j})\ket{m}\ .
\end{equation}
The following commutation relations hold, which can be shown in the same way
as for the XXX case~\cite{Skly92}:
\begin{align}
  \label{comm1}
  \comm{x_{j}}{x_{k}}&=0\ ,\\
  \label{comm2}
  X_{j}^{\pm}x_{k}&=(x_{k}\pm\eta\delta_{jk})X_{j}^{\pm}\ ,\\
  \label{comm3}
  \comm{X_{j}^{\pm}}{X_{k}^{\pm}}&=0\ ,\\
  \label{comm4}
  \comm{X_{j}^{+}}{X_{k}^{-}}&=0\quad\text{for}\quad j\neq k\ ,\\
  \label{comm5}
  X_{j}^{\pm}X_{j}^{\mp}&=\Delta(x_{j}\pm\tfrac{\eta}{2})\ ,
\end{align}
where $\Delta(\lambda)$ is the quantum determinant of the twisted monodromy
matrix
\begin{equation}
  \begin{split}
    \Delta(\lambda)&=A(\lambda+\tfrac{\eta}{2})D(\lambda-\tfrac{\eta}{2})-B(\lambda+\tfrac{\eta}{2})C(\lambda-\tfrac{\eta}{2})\\
    &=-\Bigl[\prod_{j=1}^{L}\sh(\lambda-\delta_{j}+\eta)\sh(\lambda-\delta_{j}-\eta)\Bigr]\id\ .
  \end{split}
\end{equation}
These commutation relations largely fix the action of the conjugated momenta
$X_{j}^{\pm}$ on the eigenvectors of the operators $x_{j}$. The remaining
freedom is due to the fact that the $x_{j}$-eigenvectors are determined only
up to phase factors~\cite[Thm.~3.4]{Skly92}. (In the functional language
changing these phase factors corresponds to multiplying all functions with an
arbitrary function which has no zeroes.) We will now reconstruct the
conjugated momenta from the commutation relations. This will allow us to
formulate the spectral problem for the twisted transfer matrix in the
$x_{j}$-eigenbasis without knowing the change of bases explicitly. Let
$\omega$ be the function $\omega\equiv 1$ on $X$, and define the functions
$\Delta_{j}^{\pm}$ by
\begin{equation}
  \Delta_{j}^{\pm}(x)=\bigl(X_{j}^{\pm}\omega\bigr)(x)\ .
\end{equation}
Equation~\eqref{comm2} implies
\begin{equation}
  \bigl(X_{j}^{\pm}f\bigr)(x)=\Delta_{j}^{\pm}(x)f(E_{j}^{\pm}x)
\end{equation}
for any function $f$ on $X$. Here we have introduced the shift operators
$E_{j}^{\pm}$
\begin{equation}
  E_{j}^{\pm}\colon(x_{1},\ldots,x_{j},\ldots,x_{L})\longmapsto(x_{1},\ldots,x_{j}\pm\eta,\ldots,x_{L})\ .
\end{equation}
As the functions $f$ are defined only on $X$,
\begin{equation}
  \label{Deltaboundary}
  \Delta_{j}^{\pm}(x)=0\qquad\text{whenever}\qquad E_{j}^{\pm}x\notin X
\end{equation}
must hold. The commutation relations~\eqref{comm3}--\eqref{comm5} translate
into the following conditions on the functions $\Delta_{j}^{\pm}$:
\begin{align}
  \label{Delta3}
  \Delta_{j}^{\pm}(x)\Delta_{k}^{\pm}(E_{j}^{\pm}x)&=\Delta_{k}^{\pm}(x)\Delta_{j}^{\pm}(E_{k}^{\pm}x)\ ,\\
  \label{Delta4}
  \Delta_{j}^{+}(x)\Delta_{k}^{-}(E_{j}^{+}x)&=\Delta_{k}^{-}(x)\Delta_{j}^{+}(E_{k}^{-}x)\quad\text{for}\quad j\neq k\ ,\\ 
  \label{Delta5}
  \Delta_{j}^{\pm}(x)\Delta_{j}^{\mp}(E_{j}^{\pm}x)&=\Delta(x_{j}\pm\tfrac{\eta}{2})\ .
\end{align}
We make the following ansatz:
Let the functions $\Delta_{j}^{\pm}$ be defined as
\begin{equation}
  \label{sol1}
  \Delta_{j}^{\pm}(x)=\Delta_{\pm}(x_j)\ ,
\end{equation}
where
\begin{equation}
  \label{sol2}
  \Delta_{\pm}(\lambda)=\xi_{\pm}\sh(\lambda-\delta_{1}\mp\tfrac{\eta}{2})\cdots\sh(\lambda-\delta_{L}\mp\tfrac{\eta}{2})\ .
\end{equation} 
In this definition, the constants $\xi_{\pm}$ are an arbitrary factorization of
the determinant of the twist matrix
\begin{equation}
  \label{sol3}
  \xi_{+}\xi_{-}=\det(K)=-1\ ,
\end{equation}
and the functions $\Delta_{\pm}$ factorize the quantum determinant of the
monodromy matrix in the following sense:
\begin{equation}
  \label{qdetfact}
  \Delta_{+}(\delta_{j}-\tfrac{\eta}{2})\Delta_{-}(\delta_{j}+\tfrac{\eta}{2})=\Delta(\delta_{j})\ .
\end{equation}
This ansatz satisfies the conditions given in
Eqs.~\eqref{Deltaboundary}--\eqref{Delta5}.

We now return to the spectral problem for the twisted transfer matrix
\begin{equation}
  t(\lambda)q(x)=\Lambda(\lambda)q(x)\ .
\end{equation}
The eigenvector $q$ is in our representation a complex-valued function on
$X$. We substitute $\lambda=x_{j}$ from the left and obtain
\begin{equation}
  \label{septq}
  \Lambda(x_{j})q(x)=\Delta_{+}(x_{j})q(E_{j}^{+}x)+\Delta_{-}(x_{j})q(E_{j}^{-}x)\ .
\end{equation}
The coefficients $\Delta_{\pm}$ now depend on only one coordinate
$x_{j}$. This is due to the particular solution $\Delta_{j}^{\pm}$ to the
conditions~\eqref{Deltaboundary}--\eqref{Delta5} chosen in
Eqs.~\eqref{sol1}--\eqref{sol3}, for a generic solution it would not have been
the case. The last equation~\eqref{septq} suggests the separation of variables
ansatz
\begin{equation}
  q(x)=q_{L}(x_{L})\cdots q_{1}(x_{1})\ .
\end{equation}
It remains to solve a set of one-dimensional problems
\begin{equation}
  \label{sepvar}
  \Lambda(x_{j})q_{j}(x_{j})=\Delta_{+}(x_{j})q_{j}(x_{j}+\eta)+\Delta_{-}(x_{j})q_{j}(x_{j}-\eta)\ ,\qquad j=1,\ldots,L\ ,
\end{equation}
which we recognize as the \TQ equation~\eqref{tqe}, evaluated on the discrete
lattice $X$.

Recalling that $x_{j}$ takes the values $\delta_{j}\pm\eta/2$ and using
$\Delta_{\pm}(\delta_{j}\pm\eta/2)=0$, we see that the finite-difference
equation~\eqref{sepvar} takes the form of a homogeneous system of linear
equations
\begin{equation}
  \label{sle}
  \begin{pmatrix}
    \Lambda(\delta_{j}+\tfrac{\eta}{2}) & -\Delta_{-}(\delta_{j}+\tfrac{\eta}{2})\\
    -\Delta_{+}(\delta_{j}-\tfrac{\eta}{2}) & \Lambda(\delta_{j}-\tfrac{\eta}{2})
  \end{pmatrix}
  \begin{pmatrix}
    q_{j}(\delta_{j}+\tfrac{\eta}{2})\\
    q_{j}(\delta_{j}-\tfrac{\eta}{2})
  \end{pmatrix}
  =0\ .
\end{equation}
For a nontrivial solution its determinant has to vanish
\begin{equation}
  \label{saec}
  \Lambda(\delta_{j}+\tfrac{\eta}{2})\Lambda(\delta_{j}-\tfrac{\eta}{2})=\Delta(\delta_{j})\ ,\qquad j=1,\ldots,L
\end{equation}
(we have used equation \eqref{qdetfact}). The eigenvalue $\Lambda(\lambda)$ is of
the form
\begin{equation}
  \Lambda(\lambda)=\Lambda_{-L+1}\e^{(-L+1)\lambda}+\Lambda_{-L+3}\e^{(-L+3)\lambda}+\cdots+\Lambda_{L-1}\e^{(L-1)\lambda}\ .
\end{equation}
The $L$ coefficients $\Lambda_{j}$ have to be determined from the $L$
equations~\eqref{saec}. Each of these equations defines a quadratic form in
the $L$-dimensional complex space of the coefficients of $\Lambda(\lambda)$
(see also \cite{FrSW08}). These quadratic forms intersect at
$2^{L}$ points, which correspond to the $2^{L}$ eigenvalues
$\Lambda(\lambda)$.
(Note that the eigenvalues of the transfer matrix are non-degenerate for the
anti-diagonal twist.  This explains why the method of separation of variables
does not suffer the so-called `completeness problem' of the algebraic Bethe
ansatz.  In the case of diagonal twist matrices the method cannot be applied
since the operator $B(\lambda)$ has only $L-1$ zeroes.)  

Finally, we want to remove the inhomogeneities $\delta_{j}$, which were not
present in the original problem related to the spin chain (\ref{hamil}).
Simply putting them all to zero is not possible since then the
equations~\eqref{saec} are no longer independent.  Instead, we consider the
equation
\begin{equation}
  \Lambda(\delta+\tfrac{\eta}{2})\Lambda(\delta-\tfrac{\eta}{2})
    =-\sinh^{L}(\delta-\eta)\sinh^{L}(\delta+\eta) 
\end{equation}
and its derivatives w.r.t. $\delta$ up to order $L-1$, evaluated at
$\delta=0$.  For small systems we have verified that this procedure does give
the correct eigenvalues $\Lambda(\lambda)$.

\section{\label{sec:fss}%
Finite-size scaling of the ground-state energy}
%
\subsection{Solution of the Bethe equations}
As an application of the solution of the spectral problem we now study the
description of the ground-state in terms of the solutions to the Bethe
equations.  We start by solving these equations (\ref{bae}) for the complex
Bethe rapidities $\{\lambda_{j}\}_{j=1}^{L}$ numerically for small systems.

In the antiferromagnetic massive regime, described by real $\eta$, the Bethe
rapidities of the ground-states, and only those, are purely imaginary for
small system sizes and small values of $\eta$.  As the system size grows or
for larger values of $\eta$ two of the rapidities in the ground-state form a
single `2-string' of rapidities symmetric to the imaginary axis,
$\lambda\pm\eta/2$.  Computing the energy of this state one obtains a system
size independent contribution.  This is the energy of a domain wall,
corresponding to the interfacial tension in the six-vertex model computed in
Ref.~\onlinecite{BBOY95}.

\begin{figure}[t]
  \includegraphics{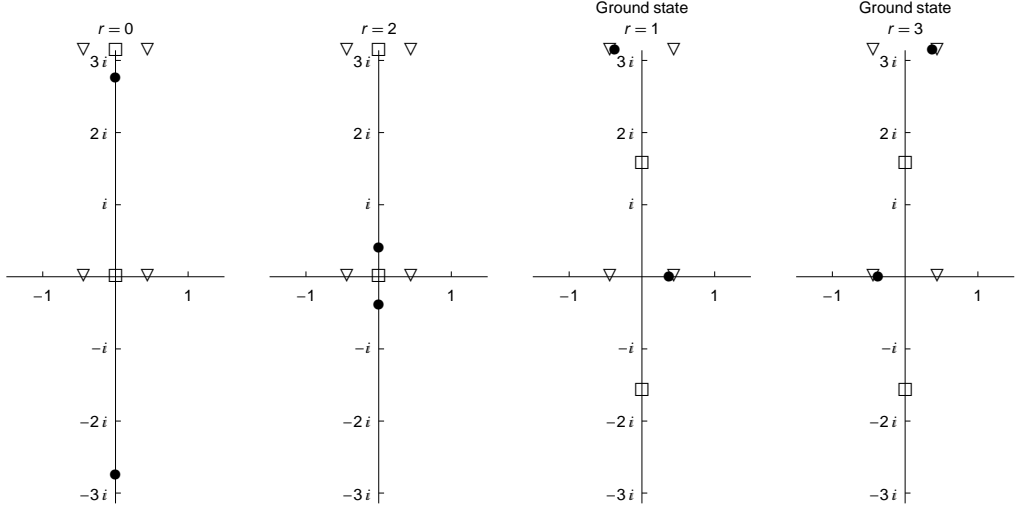}
  \caption{\label{fig:ban2ml} Distribution of the Bethe rapidities
    $\lambda_{j}$ ($\bullet$) and the rapidities ${\lambda^{(1)}}_{j}$
    ($\square$) and ${\lambda^{(2)}}_{j}$ ($\triangledown$) in the massless
    regime with $\eta=\I\pi/4$ for two lattice sites.} 
\end{figure}

\begin{figure}[t]
	\includegraphics{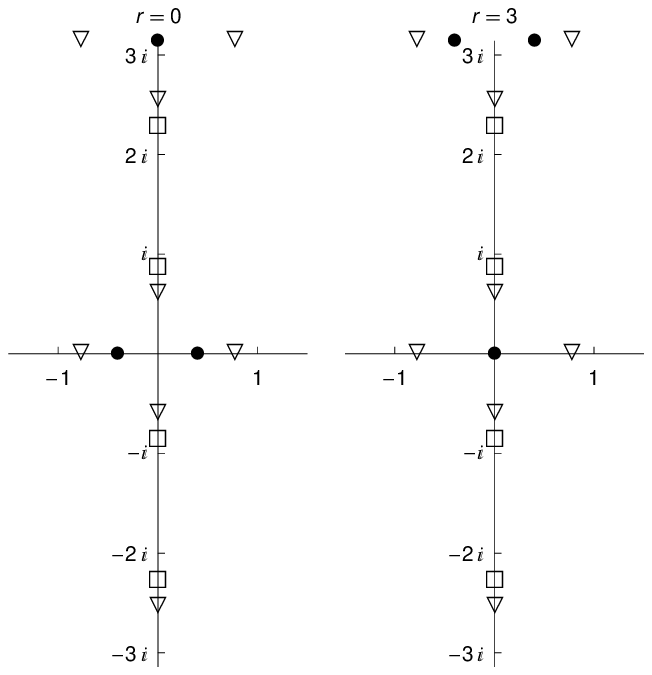}
	\includegraphics{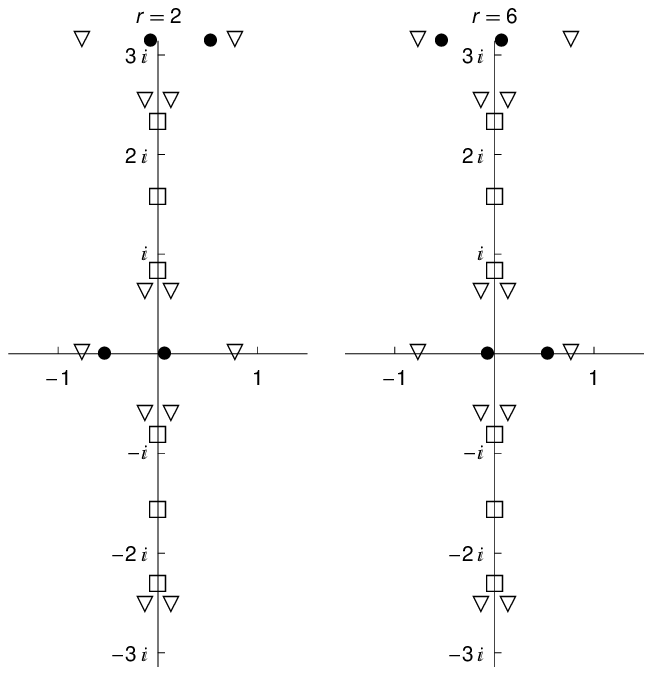}
	
	(a) $L=3$, $\eta=\I\pi/4$ \hspace{10em}
	(b) $L=4$, $\eta=\I\pi/4$
  \caption{\label{fig:ban3} Distribution of the Bethe rapidities $\lambda_{j}$
    ($\bullet$) and the rapidities ${\lambda^{(1)}}_{j}$ ($\square$) and
    ${\lambda^{(2)}}_{j}$ ($\triangledown$) for the ground-state in the
    massless regime with $\eta=\I\pi/4$ for (a) three lattice sites and (b) four lattice sites.} 
\end{figure}

Here we concentrate on the massless case, i.e. $\eta=\I\gamma$ with real
$\gamma$.  In Figures~\ref{fig:ban2ml} and \ref{fig:ban3} some results for up to 4
lattice sites and $\eta=\I\pi/4$ are shown.  As a consequence of the
periodicity of the Bethe equations the set $\{\lambda_{j}+\I\pi\}_{j=1}^{L}$
is also a solution which parametrizes a second state corresponding to a
different eigenvalue (\ref{tqe}) of the transfer matrix but has the same
energy (\ref{baenergy}).  As the lattice size increases, we find that all
Bethe rapidities of the two ground-states have imaginary parts $0$ or $\pi$.
Below we will use this observation to study the finite size scaling behaviour
of the ground-state energy.
For excited states we have not identified such a pattern which would be
necessary for a systematic analysis of the excitation spectrum starting from
the Bethe equations (\ref{bae}).

Also shown in Figures~\ref{fig:ban2ml} and \ref{fig:ban3} are the
corresponding solutions to Galleas' equations (\ref{galle}), i.e.\ the two
sets of rapidities $\{\lambda_{j}^{(1)}\}_{j=1}^{L-1}$,
$\{\lambda_{j}^{(2)}\}_{j=1}^{2(L-1)}$ and the quantum number $r$.  Galleas'
equations are invariant under separate shifts of the two sets by $\I\pi$,
therefore each rapidity appears twice and $\lambda$ and $\lambda+\I\pi$ have
to be identified.  In this parametrization, the pairs of degenerate solutions
are given by the same rapidities $\lambda_{j}^{(1)}$ and $\lambda_{j}^{(2)}$,
but their values of $r$ differ by $L$. 
Again, the distribution of the roots of (\ref{galle}) in the complex plane
does not appear to follow a simple scheme as the system size increases, not
even for the the ground-state.  Therefore, this approach is of limited use
only to study the spectrum of long chains.

To proceed, we parametrize Bethe roots corresponding to the ground-states as
follows: let $\mu_{j}$, $j=1,\ldots,m$ be the real Bethe rapidities and
$\nu_{j}$, $j=1,\ldots,n$ the real parts of the Bethe rapidities with
imaginary part $\pi$.  For even system sizes $L$ we have $m=n=L/2$, for odd
system sizes we have $m=(L\mp1)/2$, $n=(L\pm1)/2$ for two ground-state
configurations. We take the logarithm of the Bethe equations (\ref{bae}) and
obtain
\begin{equation}
  \label{logbae}
  \begin{aligned}
    L\Phi(\mu_{j})&=2\pi I_{j}+\sum_{k=1}^{m}\Theta(\mu_{j}-\mu_{k})+\sum_{k=1}^{n}\widetilde{\Theta}(\mu_{j}-\nu_{k})\ ,\qquad j=1,\dots,m\ ,\\
    L\Phi(\nu_{k})&=2\pi J_{k}+\sum_{\ell=1}^{m}\widetilde{\Theta}(\nu_{k}-\mu_{\ell})+\sum_{\ell=1}^{n}\Theta(\nu_{k}-          \nu_{\ell})\ ,\qquad k=1,\dots,n\ ,
  \end{aligned}
\end{equation}
where
\begin{align}
  \Phi(\lambda)&=-\I\ln\Bigl[\frac{\sh(\lambda-\tfrac{\I\gamma}{2})}{\sh(\lambda+\tfrac{\I\gamma}{2})}\Bigr]=\pi+2\arctan[\tnh(\lambda)\cot(\tfrac{\gamma}{2})]\ ,\nonumber\\
  \Theta(\lambda)&=-\I\ln\Bigl[\frac{\sh[\tfrac{1}{2}(\lambda-\I\gamma)]}{\sh[\tfrac{1}{2}(\lambda+\I\gamma)]}\Bigr]=\pi+ 2\arctan[\tnh(\tfrac{\lambda}{2})\cot(\tfrac{\gamma}{2})]\ ,\nonumber\\
  \widetilde{\Theta}(\lambda)&=-\I\ln\Bigl[\frac{\sh[\tfrac{1}{2}(\lambda+\I\pi-\I\gamma)]}{\sh[\tfrac{1}{2}(\lambda+     \I\pi+\I\gamma)]}\Bigr]=2\arctan[\tnh(\tfrac{\lambda}{2})\cot(\tfrac{\gamma}{2}+\tfrac{\pi}{2})]\ .\nonumber
\end{align}
We have chosen the branch of the logarithm in such a way that the functions
$\Phi$, $\Theta$ and $\widetilde{\Theta}$ are monotonically increasing. Here
$\arctan$ denotes the principal branch of the inverse tangent taking values
between $-\pi/2$ and $\pi/2$.  Our numerical solutions for small chains show
that the Bethe integers $I_{j}$ and $J_{k}$ in (\ref{logbae}) follow a simple
pattern: for even system sizes, one ground-state is obtained with
\begin{equation}
  \label{counting1}
\begin{aligned}
  I_j &= j-1 & \text{ for } j=1,\dots,\frac{L}{2} \\
  J_k &= k & \text{ for } k=1,\dots,\frac{L}{2}
\end{aligned}
\end{equation}
while for odd $L$
\begin{equation}
  \label{counting3}
\begin{aligned}
  I_j &= j-1 & \text{ for } j=1,\dots,\frac{L+1}{2} \\
  J_k &= k & \text{ for } k=1,\dots,\frac{L-1}{2}
\end{aligned}
\end{equation}
The second ground-state is obtained by interchanging the sets $\{I_j\}$ and
$\{J_k\}$.

In the thermodynamic limit we can solve the logarithmic Bethe
equations~\eqref{logbae} analytically.  Rewriting them as
\begin{equation}
  Y(\mu_{j})=I_{j}\ ,\qquad Z(\nu_{j})=J_{j}\ 
\end{equation}
with the `counting functions' $Y$ and $Z$ which satisfy
$Y(-\infty)=Z(-\infty)=0$ and $Y(+\infty)=m$, $Z(+\infty)=n$.  Assuming that
the distributions of the rapidities $\mu_{j}$ and $\nu_{j}$ can be described
by countinuous densities $\rho$ and $\sigma$ in the thermodynamic limit they
are given by the derivatives of the counting functions
\begin{equation}
	\rho(\lambda)=\frac{1}{L}Y'(\lambda)\ ,\qquad\sigma(\lambda)=\frac{1}{L}Z'(\lambda)\ .
\end{equation}
The logarithmic Bethe equations become a pair of coupled integral equations
\begin{equation}
  \begin{aligned}
    2\pi\rho(\lambda)&=\Phi'(\lambda)-\int\dv\lambda'\,\Theta'(\lambda-\lambda')\rho(\lambda')-\int\dv\lambda'\,     \widetilde{\Theta}'(\lambda-\lambda')\sigma(\lambda')\ ,\\
    2\pi\sigma(\lambda)&=\Phi'(\lambda)-\int\dv\lambda'\,\widetilde{\Theta}'(\lambda-\lambda')\rho(\lambda')-\int\dv\lambda'\,\Theta'(\lambda-\lambda')\sigma(\lambda')\ 
  \end{aligned}
\end{equation}
which can be solved by Fourier transform, yielding
\begin{equation}
  \rho(\lambda)=\sigma(\lambda)=\frac{1}{2\gamma}\frac{1}{\ch(\lambda\pi/\gamma)}\ .
\end{equation}
Here we have assumed without loss of generality $0<\gamma<\pi$. For the ground-state 
energy per lattice site in the thermodynamic limit we find with
(\ref{baenergy}) 
\begin{equation}
  \label{gse}
  \varepsilon_{\infty}=\cos\gamma
   -\frac{\sin\gamma}{\gamma}\int\frac{\dv\lambda}{\ch(\lambda\pi/\gamma)}\,
   \frac{2\sin\gamma}{\ch(2\lambda)-\cos\gamma}\ .
\end{equation}
This agrees with the result for the untwisted chain \cite{YaYa66b}, though the
distribution of the Bethe rapidities is different.

\subsection{Conformal field theory}
The low energy effective field theory for the XXZ model in the massless regime,
$-1<\cos \gamma\le1$, is well known to be that of a free boson with
compactification radius $\sqrt{2\pi}R=\sqrt{(\pi-\gamma)/\pi}$.  Conformal field
theory predicts the finite size scaling of the energies of the ground-state
and the low lying excitations \cite{Affl86,BlCN86} for a system with periodic
boundary conditions as
\begin{equation}
\label{gsc}
  E_{h\bar{h}}(L) = L\,e_\infty - \frac{\pi v_F}{6L}\,c 
                      + \frac{2\pi v_F}{L}\,(h+\bar{h}) + o(L^{-1})
\end{equation}
where $e_{\infty}$ is the energy density in the ground-state, which is in our
case given by Eq.~\eqref{gse}, and $v_F = \frac{2\pi}{\gamma}\,\sin\gamma$ is
the `Fermi' velocity of elementary excitations in the system.  The universal
number $c$ is the central charge of the conformal field theory, for the free
boson it is $c=1$.

{}From the energies appearing in the spectrum (\ref{gsc}) for a given lattice
realization of the field theory one can identify the operator content of the
latter: $h$ and $\bar{h}$ are the conformal weights of primary operators of
the CFT.  By choosing particular boundary conditions one obtains different
sectors of the theory (i.e.\ certain representations of the global symmetry
group $O(2)$ of the model (\ref{hamil})).
For diagonal boundary conditions (\ref{Kmatrices}) with $\phi=0$ and $\pi/2$
the symmetry of the bulk Hamiltonian (\ref{hamil}) is preserved and the
spectrum is given in terms of the highest weights of two commuting $U(1)$
Kac-Moody algebras.  The scaling dimensions of the primary operators ${\cal
  O}_{n,m}$ are given in terms of the eigenvalues $n$ of the $U(1)$ charge
operator $\frac{1}{2}\sum_j\sigma_j^z$ and momentum $2\pi m/L$ by
\begin{equation}
\label{cdim}
  x_{n,m} = h_{n,m}+\bar{h}_{n,m} = \pi R^2n^2  + \frac{m^2}{4\pi R^2}
\end{equation}
($m$ takes integer (half odd integer) values for $\phi=0$ ($\phi=\pi/2$)).
A diagonal twist (\ref{Kmatrices}) with angle $\phi\ne0,\pi/2$ breaks the
global symmetry to $SO(2)$.  The dimensions of primary operators are
$x_{n,m+\phi/\pi}$.  As a consequence the finite size scaling of the lowest
energy state energy is changed to $E_\mathrm{GS}(L) = L\,e_\infty - {\pi
  v_F}\,c_\mathrm{eff}/{6L}$ with an effective central charge \cite{AlBB88}
\begin{equation}
\label{ceff-per}
  c_\mathrm{eff} = 1-12\,x_{0,\phi/\pi} =
  1-\frac{6\phi^2}{\pi(\pi-\gamma)}\,,\quad |\phi|\le\frac{\pi}{2}.
\end{equation}

In the case of anti-diagonal twisted boundary conditions the $O(2)$ bulk
symmetry is broken to $Z_2\otimes Z_2$ with the factors being generated by
rotation around the $z$-axis by $\pi$ and a global spin flip, respectively.
The low-energy spectrum is that of a $U(1)$-twisted Kac-Moody algebra without
conserved charge.  In this case the conformal weights are \cite{ABGR88}
\begin{equation}
\label{cw-tw}
  (h,\bar{h})_{k_1,k_2} = \left(\frac{(4k_1+1)^2}{16},\frac{(4k_2+1)^2}{16}
  \right)  
\end{equation}
with integer $k_i$.

We have solved the Bethe equations \eqref{logbae} for the ground-state of the
spin chain numerically for system sizes up to $L=500$. From (\ref{cw-tw}) we
expect the lowest energy state to be that with conformal weights
$(h,\bar{h})_{0,0} = (\frac{1}{16},\frac{1}{16})$ for both even and odd number
of lattice sites.  With (\ref{gsc}) this leads to the CFT prediction for the
finite size scaling of the energy
\begin{equation}
\label{gsctw}
  E_\mathrm{GS}(L) = L\epsilon_\infty + \frac{\pi v_F}{6L}
  \left(-\frac{1}{2}+12\left(\frac{1}{16}+\frac{1}{16}\right)+ R(L)\right)\,,
  \qquad \lim_{L\to\infty} R(L) = 0.
\end{equation}
The resulting effective central charge $c_\mathrm{eff} = -1/2$ is independent
of the anisotropy which agrees nicely with our numerical data presented in
Table~\ref{tab:fse}.  In the isotropic limit of the XXZ model, $\gamma\to0$,
the spectrum depends only on the eigenvalues of the twist matrix
(\ref{twist-matrix}) and therefore the anti-diagonal twist is unitary
equivalent to a diagonal one with twist-angle $\pi$ (\ref{ceff-per}).  

The corrections $R(L)$ to the scaling (\ref{gsctw}) are a consequence of the
fact that the lattice hamiltonian (\ref{hamil}) differs from the conformally
invariant hamiltonian of the continuum theory by terms involving irrelevant
operators \cite{Card86a}.  Perturbation of the conformal theory with an
irrelevant operator with scaling dimension $x>2$ leads to $R(L)\propto
L^{2-x}$.  Therefore, by analyzing these corrections in the numerical data one
can identify the leading irrelevant perturbation of the lattice hamiltonian.

\begin{figure}[t]
\includegraphics[width=0.8\textwidth]{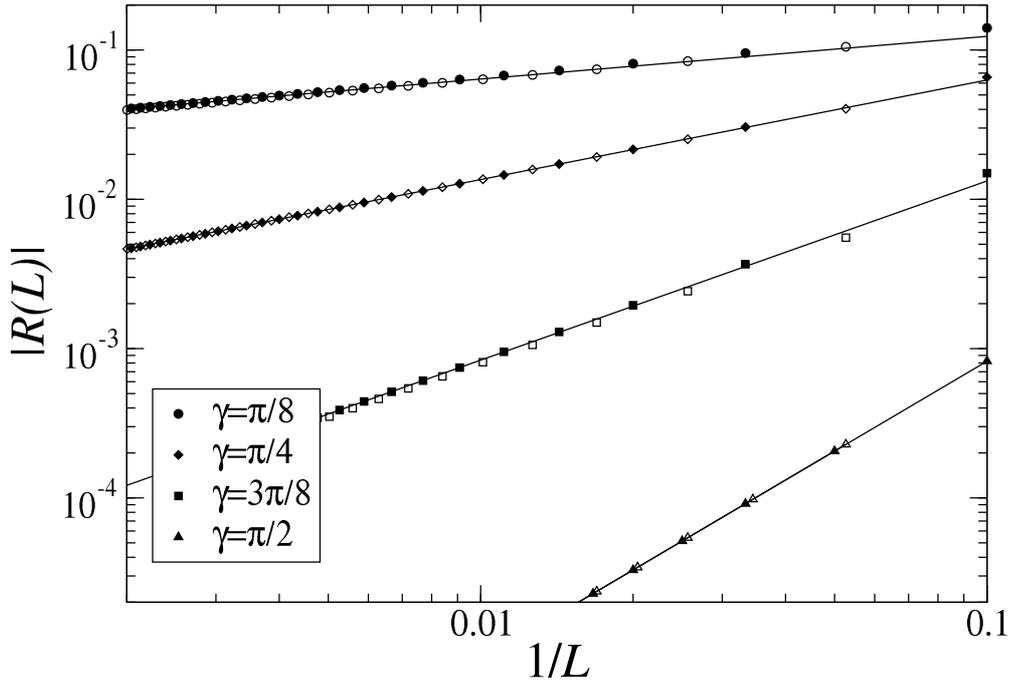}
\caption{\label{fig:scal}Corrections to the scaling of the ground-state
  energy: Closed (open) symbols are numerical finite size data for even (odd)
  system sizes, lines are fits to a power law $R(L)\propto
  L^{-2\gamma/(\pi-\gamma)}$.}
\end{figure}

For the periodic XXZ chain and the model with diagonal twist the corrections
to scaling vanish with an exponent $x-2=\max(4\gamma/(\pi-\gamma),2)$
\cite{WoEc87a,AlBB88}.  The hamiltonian of the XXZ model with anti-periodic
twisted boundary conditions, however, is related to the thermal operator of
the Ashkin-Teller model and the leading corrections to scaling $R(L)$ are
generated by the operator ${\cal O}_{0,2}$ \cite{AlBB87}, i.e.\
\begin{equation}
\label{scalcorr}
  R(L) \propto L^{2-x_{0,2}}  = L^{-\frac{2\gamma}{\pi-\gamma}}\,.
\end{equation}
As shown in Figure~\ref{fig:scal}, this provides an excellent fit for our
numerical data.  The dependence (\ref{scalcorr}) on the anisotropy parameter
explains the slow convergence towards $c_\mathrm{eff}$ in
Table~\ref{tab:fse} as $\gamma\to0$.

\begin{table}[t]
  \caption{\label{tab:fse}
    Finite size scaling of the ground-state energy in
    systems with even (odd) number of sites for
    several values of the anisotropy $\gamma$:  Shown are the numerical
    results for $-c_{\mathrm{eff}}= (6/\pi v_F)\, L(E_0(L)-L\,e_\infty)$ for
    lattices of size $L$ together with the extrapolation to $L=\infty$.
  }
\begin{ruledtabular}
\begin{tabular}{rllll}
L & $\gamma=\pi/8$ & $\gamma=\pi/4$ & $\gamma=3\pi/8$ & $\gamma=\pi/2$ \\\hline
30  & 0.59536849 & 0.53046883 & 0.50367226 & 0.50009141 \\
60  & 0.57644644 & 0.51909229 & 0.50156060 & 0.50002285 \\
120 & 0.56176871 & 0.51200578 & 0.50067152 & 0.50000571 \\
240 & 0.55014811 & 0.50755893 & 0.50029553 & 0.50000143 \\
480 & 0.54084022 & 0.50476118 & 0.50014993 & 0.50000035 \\
extr. & 0.51(2)  & 0.500(1)   & 0.5000(1)  & 0.500000(1)\\
%
\hline
29  & 0.40781510 & 0.46933409 & 0.49658570 & 0.50009782 \\
59  & 0.42572231 & 0.48079424 & 0.49850626 & 0.50002363 \\
119 & 0.43973453 & 0.48794414 & 0.49934855 & 0.50000494 \\
239 & 0.45088912 & 0.49242107 & 0.49972106 & 0.50000144 \\
479 & 0.45987007 & 0.49523121 & 0.49990010 & 0.50000036 \\
extr. & 0.48(2)  & 0.499(1)   & 0.4999(1)  & 0.500000(1)
\end{tabular}
\end{ruledtabular}
\end{table}
\section{Summary and Conclusion}
For the XXZ spin chain with anti-diagonal twist which does not allow for a
solution of the spectral problem by means of the algebraic Bethe ansatz due to
the lack of a reference state we have derived the functional equations
(\ref{tqe}), originally obtained using Baxter's method of commuting transfer
matrices, employing different methods: restricting the anisotropy to roots of
unity $\eta=\I\pi/(p+1)$ the \TQ equation of \cite{BBOY95} was obtained
by truncation of the fusion hierarchy.  In a second approach, namely through
Sklyanin's separation of variables, we have used a representation of the
Yang-Baxter algebra on a space of symmetric functions defined on a discrete
lattice.  In this formulation the spectral problem could again be recast in
the form of the same \TQ equation (\ref{sepvar}) -- in this case, however, it
has to be solved on the lattice of singular points of this functional equation
only.  This is unlike the situation for the spin chain with open boundaries
and non-diagonal boundary fields where different functional equations have
been found within different approaches.

In the lattice formulation of the \TQ equation arising from the separation of
variables the computation of the eigenvalues amounts to finding roots of $L$
coupled polynomial equations (\ref{saec}), therefore this approach appears to
be best suited to determine the spectrum of small systems.  A similar
limitation holds for the recent approach of Galleas \cite{Galleas08}.  For an
efficient solution of the \TQ equation in the thermodynamic limit the
parametrization (\ref{Q-func0}) or (\ref{Q-Func}) of the solution to the
functional equations has to be used which leads to $L$ algebraic Bethe
equations (\ref{bae}).  At least for the ground-state energy of the spin chain
these equations can be solved, e.g. to obtain the interfacial tension of the
six-vertex model (see \cite{BBOY95}) or to identify the operator
content of the low energy effective theory in the massless regime.

Note, that together with the eigenvalues $\Lambda(\lambda)$ one obtains the
functions $q(\lambda)$ from the \TQ equation in either approach.  It is quite
clear from the separation of variables that these functions contain the
complete information on the eigenstates of the transfer matrix.  However,
unlike the situation with the algebraic Bethe ansatz, where one has an
expression for the eigenstates in terms of the generators of the Yang-Baxter
algebra, the explicit transformation from the $q$-functions to state vectors
in the Hilbert space of the spin chain is not known.
\begin{acknowledgments}
  We thank A. Osterloh and A. Seel for helpful discussions.  This work has
  been supported by the Deutsche Forschungsgemeinschaft under grant
  no.\ Fr~737/6.
  SN acknowledges funding by the FWF and by the EU (SCALA, OLAQUI, QICS).
\end{acknowledgments}
\enlargethispage{\baselineskip}
%
%
%

\end{document}